\documentstyle[aps]{revtex}

\begin{document}
\title{Quantum memory for individual polarization photon }
\author{Guo-Ping Guo\thanks{%
Electronic address: harryguo@mail.ustc.edu.cn },Guang-Can Guo}
\address{Key Laboratory of Quantum Information, University of Science and Technology\\
of China, Chinese Academy of Science, Hefei, Anhui, P. P. China, 230026}
\maketitle

\begin{abstract}
\baselineskip12ptHere we present an experimentally feasible quantum memory
for individual polarization photon with long-lived atomic ensembles
excitations. Based a process similar to teleportation, the memory is
reversible. And the storage information can be effortlessly read out and
transferred back to photon. Although it successes with only a probability of
1/4, it is expected valuable in various quantum information processing,
especially those cases where polarized photons are employed. The physical
requirements are moderate and fit the presest technique.

PACS number(s): 03.65.Ud, 03.67.-a, 42.50.Gy. 42.50.-p
\end{abstract}

\baselineskip12pt

Quantum information theory has been rapidly developed in recent years for
its promise to the absolutely secure transmission of secret messages and the
faithful transfer of unknown quantum states. Now it has been realized that
quantum memory is very valuable, or even absolutely necessary in various
quantum information applications such as quantum dense coding\cite{ds},
quantum secret sharing\cite{qss1,qss2,g,ss,s5}, quantum data hiding\cite
{qdh,qdhg}, and long distance quantum communication\cite{du}. The employment
of quantum memory can also greatly improve some quantum key distribution
schemes efficiencies\cite{ops}. For example, the BB84 protocol can be
modified to be 100\% efficient, when we can take a procedure of making
measurements after the announcement of bases\cite{bb}.

As photonic channel appears to be the most attractive candidate for the
physical implementation of quantum communication, a quantum memory for
individual polarization photon will be significance. However most previous
quantum memory protocols need to employ uncommon High $Q$ cavity\cite{hq}.
Recently, it has been noted that the emission of a photon in the forward
direction is correlated with an atomic ensemble symmetric mode excitation%
\cite{du}. In paper\cite{gg}, we have proposed a quantum memory protocol to
store individual polarization photon with atomic ensembles. The protocol is
argued as one application of a novel state combining the atomic ensemble
collective states and individual photon polarization states. Basically due
to the depolarization from collisions, the storage time in the atomic
ensemble collective states can be about $10ms$ to $1s$. However the above
scheme is based on the Raman process. Thus to obtain an efficient
information storage, an unusual photon non-demolition measurement device is
needed, to cope with the randomness for the producing of Stokes photon.

Here we propose a reversible quantum memory scheme for individual
polarization photons, which is well based on the atomic ensemble anti-Raman
process. As there is no random for the generation of anti-Stokes photon, no
photon non-demolition measurement device is required in the present quantum
memory protocol. The quantum memory can be divided into three stages: (i)
initial preparation of the memory. In this stage, two pairs of atomic
ensembles are prepared into an effective entangled state through Raman
process. (ii) storage of a single photon polarization state conditioned to a
Bell measurement. Through a procedure similar to teleportation, the state of
the incoming photon is probably teleported to the atomic ensembles. (iii)
manipulation and read out of this state. As the atomic ensembles collective
excited modes can be effortlessly transferred to photon modes, the stored
quantum information can be efficiently manipulated and read out in the
present reversible quantum memory. Although the protocol only succeeds with
a probability of $1/4$, it can be still worthiness in many applications
where quantum memory is absolutely necessary. The physical requirements are
moderate and fit the present-day experimental techniques.

The basic elements of our system are two pairs of ensembles, A1, A2 and B1,
B2, of alkali atoms with the relevant level structure as shown in Fig. 1. A
pair of metastable low states $\left| g\right\rangle $ and $\left|
s\right\rangle $ correspond to Zeeman sublevels of the electronic ground
state of alkali-metal atoms. Its experimental realization can be either a
room-temperature atomic gas or a sample of cold trapped atoms where long
lifetimes for the relevant coherence has been both observed\cite
{dn20,d14,d15}. To facilitate enhanced coupling to light, the atomic medium
is preferably optically thick along one direction\cite{dn20,d14,d15}. This
can be achieved either by working with a pencil-shaped atomic sample or by
placing the sample in a low-finesse ring cavity\cite{dn17,dn25}.

For these ensembles, we can define an operator $S=(1/\sqrt{N_a}%
)\sum_{i=1}^{N_a}$ $\left| g\right\rangle _i\left\langle s\right| ,$ where $%
N_a\gg 1$ is the total atom number in one ensemble. All atoms are initially
prepared through optical pumping to the ground state $\left| g\right\rangle
, $ which is effectively the vacuum state $\left| 0\right\rangle _a=\otimes
_i\left| g\right\rangle _i$ of the operator $S$. Illuminated with a short
off-resonant laser pulse to induce atom Raman transition into state $\left|
s\right\rangle $, atomic ensemble will have a probability $p$ to be excited
to the symmetric collective atomic mode $S$, and result in the state $%
S^{\dagger }\left| 0\right\rangle _a=(1/\sqrt{N_a})\sum_{i=1}^{N_a}\left|
s\right\rangle _i=\left| S\right\rangle .$ Uniquely correlated with this
ensemble excited mode, a forward-scattered Stokes photon will be emitted out
from the ensemble, which is $co$-propagating with the pump light. For a
short light-atom interaction time $t_\Delta $, the probability for this
Raman process $p\ll 1$. The probability for the ensemble to have more than
one excitation is exceeding small and can be neglected.

When we re-pump an ensemble in the excitation mode $S$ with an anti-pump
pulse, the ensemble excitation mode will be transformed into an optical
excitation mode. And the excited ensemble will emit out an anti-Stokes
photon in horizontal polarization mode $a_h^{+}$ within the anti-Raman
process. Due to the collectively enhanced coherent interaction, such
transfer can be almost $100\%$ efficient, which has been demonstrated both
in theory\cite{d17} and in experiments\cite{d14,d15}. Thus there is no
random for the generation of anti-Stokes photon in mode $a_h^{+}$.

In the first stage of the present quantum memory, we firstly illuminate the
two pairs of ensembles A1, A2 and B1, B2 in ground state respectively with a
short off-resonant laser pulse to prepare them into the state $%
(S_{A1}^{+}+S_{A2}^{+})\left| 0\right\rangle _a$ and $%
(S_{B1}^{+}+S_{B2}^{+})\left| 0\right\rangle _a$. Following the discussing
of the reference\cite{du,du1}, the two ensembles A1, A2 or B1, B2 are
pencil-shaped, and illuminated by the synchronized classical pumping pulses
as shown in Fig. 2. After filtering out the pumping light with frequency
selective filter, the forward-scattered Stokes pulses are collected and
coupled to optical channels. The pulses after the transmission channels
interfere at a 50\%-50\% beam splitter (BS), with the outputs detected
respectively by two single-photon detectors D1 and D2. If there is a click
in D1 or D2, the two ensembles M1 and M2 are successfully prepared into the
state $(S_{M1}^{+}+e^{i\Phi }S_{M2}^{+})\left| 0\right\rangle _a$. Here M
represents A pair or B pair of two ensembles. And $\Phi $ denotes an unknown
but fixed difference of the phase between the two ensembles. Otherwise, we
first apply a re-pump pulse (to the transition $\left| s\right\rangle $ $%
\rightarrow \left| e\right\rangle $) to the two ensembles, to set them back
to the vacuum ground state. Then the same classical laser pulses as the
first round are applied to the ensembles to induce the transition $\left|
g\right\rangle $ $\rightarrow \left| e\right\rangle .$ The
forward-scattering Stokes pulses are similarly detected after the beam
splitter. The process is repeated until finally we have a click in the
detector D1 or D2. Then the two ensembles are prepared in the state $\left|
\Psi \right\rangle _{M1,M2}=(S_{M1}^{+}+S_{M2}^{+})\left| 0\right\rangle _a$%
, with M represents A pair or B pair of two ensembles. Without loss of
generality, we have neglected the fixed phase $\Phi $\cite{du}. Thus the
four ensembles can be written in the product state (which is unnormalized) 
\begin{equation}
(S_{A1}^{+}+S_{A2}^{+})(S_{B1}^{+}+S_{B2}^{+})\left| 0\right\rangle _a,
\end{equation}
where $\left| 0\right\rangle _a$ is the total vacuum state of the whole
system. We will see later that this state can effectively act similarly as
EPR state $\frac 1{\sqrt{2}}(S_{A1}^{+}S_{B2}^{+}+S_{A2}^{+}S_{B1}^{+})%
\left| 0\right\rangle _a$ in the present protocol. These four atomic
ensembles in the state (1) form the quantum memory for individual
polarization photon.

When the polarization photon needed recording comes, we anti-pump the
ensembles A1 and B1 in turn to transfer their excitation mode to the
anti-Stokes photon polarization mode $a_h^{\dagger }$. This
forward-scattering anti-Stokes photon and the incoming photon are then
measured with a Bell states analyzer as shown in Fig. 3. The $\lambda /2$
plate between the two ensembles is employed to rotate the polarization modes
of anti-Stokes photon emitted from the A1 ensemble, where $\lambda $ is the
wavelength of the anti-Stokes photon. Its function can be denoted by
operator $P=$ $(a_h^{\dagger }a_v+a_v^{\dagger }a_h)/\sqrt{2}$ with $h$ ($v$%
) represents the horizontal (vertical) polarization mode. The anti-Stokes
photons transferred from two ensembles A1 and B1, the residual two ensembles
A2 and B2, and the incoming photon can be easily written in the total state
(which is unnormalized) 
\begin{equation}
(a_v^{\dagger }+S_{A2}^{+})(a_h^{\dagger }+S_{B2}^{+})(\alpha a_h^{+}+\beta
a_v^{+})\left| 0\right\rangle ,
\end{equation}
where $\left| 0\right\rangle $ is the vacuum state of the total system. In
the Bell states analyzer\cite{gg1}, the polarizing beam splitter (PBS)
reflects vertical photons and transmits horizontal photons. The two $\lambda
/2$ plates in this analyzer are employed to rotate the polarization of the
Stokes, which transfers the $a_h^{+}$ mode photon into $(a_h^{+}+a_v^{+})/%
\sqrt{2}$ and $a_v^{+}$ mode photon into $(a_h^{+}-a_v^{+})/\sqrt{2}.$ Here $%
\lambda $ is again the wavelength of anti-Stokes photon. After the
operations of the Bell states analyzer, the system total state is
transformed into the following form (which is unnormalized): 
\begin{equation}
(a_{D_h^d}^{+}-a_{D_v^d}^{+}+2S_{A2}^{+})(a_{D_h^u}^{+}+a_{D_v^u}^{+}+2S_{B2}^{+})(\alpha (a_{D_h^d}^{+}+a_{D_v^d}^{+})+\beta (a_{D_h^u}^{+}-a_{D_v^u}^{+}))\left| 0\right\rangle .
\end{equation}
Here $a_{D_l^p}^{+}$ represents the photon mode before the detector $D_l^p$,
with $p=u,d$ and $l=h,v$.

It can be easily seen from the state form (3) that there is a probability of 
$1/4$ to have coincidence clicks between two detectors $D_h^u$ and $D_h^d$
or $D_v^u$ and $D_v^d$. In this case, the two detectors detect three or two
photons, and the residual two atomic ensembles A2 and B2 are correspondingly
projected into vacuum state or $(\alpha S_{A2}^{+}+\beta S_{B2}^{+})\left|
0\right\rangle $. Similarly, there is a probability of $1/4$ to have
coincidence clicks between two detectors $D_h^u$ and $D_v^d$ or $D_v^u$ and $%
D_h^d$. And then the residual two atomic ensembles A2 and B2 are projected
with an equal probability into vacuum state or $(\alpha S_{A2}^{+}-\beta
S_{B2}^{+})\left| 0\right\rangle $. When the detectors can distinguish one
photon from two, we can exclude those cases that the two ensembles A2 and B2
remain in vacuum state. Thus up to a Pauli operation, the state of the input
photon can be faithfully teleported to the two residual ensembles A2 and B2
with a total probability of $1/4$. Otherwise, post-select measurements are
needed for the present quantum memory.

Obviously, we can conveniently manipulate and readout the storage
information when we transfer the $S$ mode excitations to optical
excitations. Anti-pumping the two ensembles A2 and B2 in turn as the storage
procedure, we can almost $100\%$ efficiently transfer the quantum
information stored in these two ensembles to the produced anti-Stokes
photon. Thus the present protocol can be argued as a reversible quantum
memory. The quantum information can be probably transferred from individual
photon to atomic ensembles. After storage, the information can be faithfully
transferred back to individual photon. There are two reasons that the
present quantum memory for individual polarization photon only succeeds with
a probability of $1/4$. Firstly, the four ensembles are prepared in product
state (1), which only acts effectively as EPR state$\frac 1{\sqrt{2}}%
(S_{A1}^{+}S_{B2}^{+}+S_{A2}^{+}S_{B1}^{+})\left| 0\right\rangle _a$ with a
post selection probability of $1/2$. Secondly, we can only have in-complete
Bell measurements with linear optics. However, this probability quantum
memory for individual polarization photon can be still valuable for the
cases that quantum memory is absolutely necessary.

As the incoming state is teleported to the storage qubit in this memory
procedure, the photon wanted recording and the two storage ensembles doesn't
directly interact with each other. The fidelity of the four ensembles
product state and the precision of the Bell states measurements will
directly effect the fidelity of the present quantum memory. The memory
fidelity is defined as $F=\left| \left\langle \Phi \right| \Psi \rangle
\right| ^2$ with $\left| \Phi \right\rangle $ and $\left| \Phi \right\rangle 
$ represent the incoming photon state and the two ensembles storage state
respectively. In view of noise in the preparing of the four ensembles
product state, they are in fact in the state (unnormalized) 
\begin{equation}
\rho =(c_1\left| 0\right\rangle _{A1,A2}\left\langle 0\right| +\left| \Psi
\right\rangle _{A1,A2}\left\langle \Psi \right| )\otimes (c_2\left|
0\right\rangle _{B1,B2}\left\langle 0\right| +\left| \Psi \right\rangle
_{B1,B2}\left\langle \Psi \right| ),
\end{equation}
where the small vacuum coefficient $c_1$ and $c_2$ are determined by the
dark count rates of the photon detectors. It can be easily seen that this
modified state only effects the memory efficiency and makes the present
quantum memory need post-selection measurements. In fact, the noise caused
by the single-photon detectors dark count is very small since the dark count
probability is about $10^{-5}$ in a typical detection time window $0.1\mu s$%
. We can also safely neglect the Bell state measurements noise caused by the
single-photon detectors dark count. To make precision Bell state
measurements, we should still assume that the anti-Stokes photons and the
incoming photon interfere at the polarization beam splitter (PBS) of the
Bell states analyzer. This may be the most rigorous requirement for the
present quantum memory protocol.

In conclusion, we have proposed a reversible quantum memory to store
individual polarization photon states in the long-lived atomic ensemble
excitations, which succeeds with a probability of $1/4$. Different from the
prior schemes, the present quantum memory is based on atomic ensemble
anti-Stokes process. No uncommon single photon non-demolition measurement
device or high Q cavity is needed. In addition, the storage information can
be effortlessly manipulated, read out, and transferred back to photon. This
quantum memory involves only laser manipulation of atomic ensembles,
(polarization) beam splitter, and single-photon detectors with moderate
efficiencies. Its experiment implementation is currently in progress in our
laboratory.

This work was funded by National Fundamental Research Program(2001CB309300),
National Natural Science Foundation of China, the Innovation funds from
Chinese Academy of Sciences, and also by the outstanding Ph. D thesis award
and the CAS's talented scientist award entitled to Luming Duan.

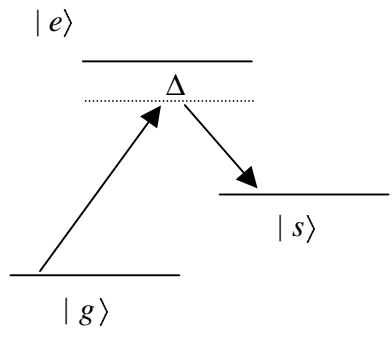
Figure 1: The relevant level structure with $\left| g\right\rangle ,$ the
ground state, $\left| e\right\rangle ,$ the excited state, and $\left|
s\right\rangle $ the metastable state for storing a qubit. The transition $%
\left| g\right\rangle \rightarrow \left| e\right\rangle $ is coupled by the
classical laser (the pump light) with Rabi frequency $\Omega ,$ and the
forward-scattered Stokes light comes from the transition $\left|
e\right\rangle \rightarrow \left| s\right\rangle ,$ which is right-handed
rotation. For convenience, we assume off-resonant coupling with a large
detuning $\Delta .$

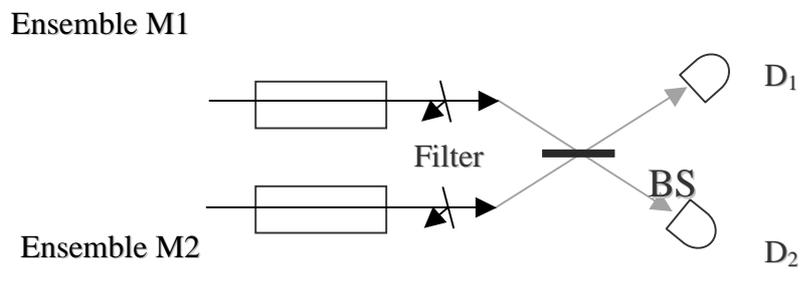
Figure 2: Illuminate the M pair of ensembles M1 and M2 with synchronized
classical pumping pulses, where $M=A,B$. The forward-scattered Stokes pulses
are collected and coupled to optical channels after the filters, which are
frequency selective to filter the pumping light. The pulses after the
transmission channels interfere at a $50\%-50\%$ beam splitter (BS), with
the outputs detected respectively by two single-photon detectors D1 and D2.

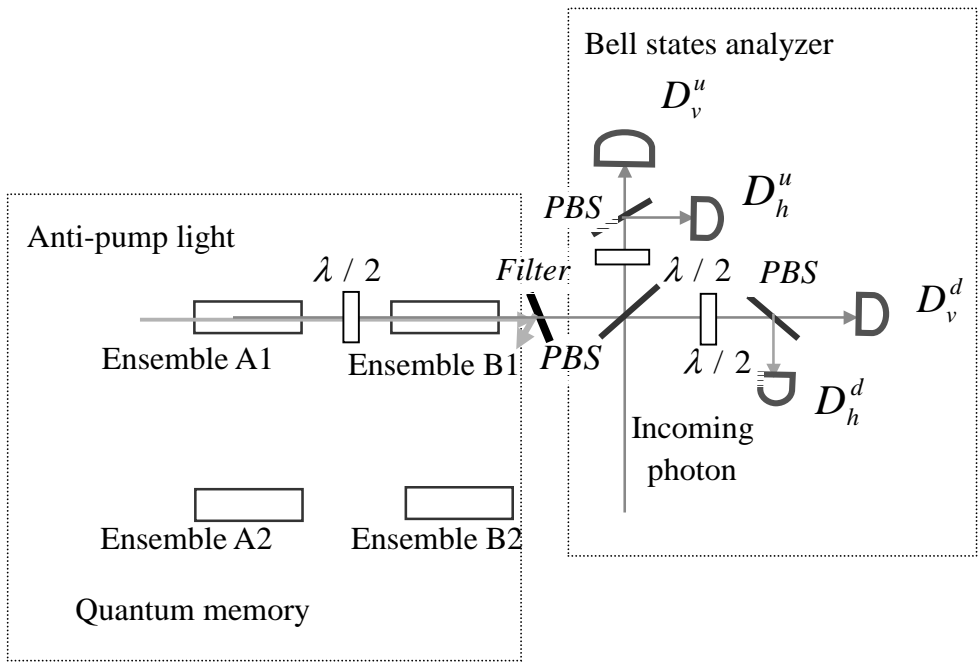
Figure 3. Schematic setup for the quantum memory procedure. The $\lambda /2$
plate between the two ensembles is employed to rotate the horizontal
polarization mode anti-Stokes photon into vertical mode. Anti-pump the
ensembles A1 and B1 in turn. After filtering out the anti-pumping light, the
forwarding-scattering anti-Stokes photon and the incoming photon needing
recording are measured with a Bell states analyzer. This analyzer can divide
the four Bell states into three class: \{$\left| \Phi \right\rangle ^{+}=%
\frac 1{\sqrt{2}}(a_h^{+}a_h^{+}+a_v^{+}a_v^{+})\left| 0\right\rangle $\}, \{%
$\left| \Phi \right\rangle ^{+}=\frac 1{\sqrt{2}}%
(a_h^{+}a_h^{+}-a_v^{+}a_v^{+})\left| 0\right\rangle $\} and \{$\left| \Psi
\right\rangle ^{\pm }=\frac 1{\sqrt{2}}(a_h^{+}a_v^{+}\pm
a_v^{+}a_h^{+})\left| 0\right\rangle $\}. The two half-wave plates in this
analyzer are employed to transfers the $a_h^{+}$ mode photon into $%
(a_h^{+}+a_v^{+})/\sqrt{2}$ and $a_v^{+}$ mode photon into $%
(a_h^{+}-a_v^{+})/\sqrt{2}$.

\end{document}